\documentclass[11pt,onecolumn,preprintnumbers]{revtex4}
\usepackage{amsmath,amsfonts,amssymb}
\usepackage[utf8]{inputenc}
\usepackage{graphicx}
\usepackage{epstopdf}
\usepackage{float,placeins}
\usepackage[caption=false]{subfig}
\usepackage[driverfallback=dvipdfm,hyperfigures,breaklinks,colorlinks,citecolor=blue,linkcolor=red]{hyperref} 
\usepackage{color}

\newcommand{\bs}{\mathbf{s}}
\newcommand{\W}[2]{W\left(#1\, \arrowvert \,#2\right)}

\def\be{\begin{equation}}
\def\ee{\end{equation}}

\begin{document}
\title{A simple analytical description of the non-stationary dynamics in Ising spin systems}

\author{Eduardo Dom\'inguez V\'azquez}
\affiliation{Henri Poincar\'e Group of Complex Systems and
  Department of Theoretical Physics, Physics Faculty, University of
  Havana, La Habana, CP 10400, Cuba.}

\author{Gino Del Ferraro}
\affiliation{Department of Computational Biology, AlbaNova University Centre, SE-106 91 Stockholm, Sweden}

\author{Federico Ricci-Tersenghi}
\affiliation{Dipartimento di Fisica, INFN-Sezione di Roma 1 and CNR-Nanotec, Universit\'a La Sapienza, Piazzale Aldo Moro 5, I-00185 Roma, Italy}

\date{\today}

\begin{abstract}
The analytical description of the dynamics in models with discrete variables (e.g.\ Ising spins) is a notoriously difficult problem, that can be tackled only under some approximation. 
Recently a novel variational approach to solve the stationary dynamical regime has been introduced by Pelizzola [Eur.~Phys.~J.~B, 86 (2013) 120], where simple closed equations are derived under mean-field approximations based on the cluster variational method. Here we propose to use the same approximation based on the cluster variational method also for the non-stationary regime, which has not been considered up to now within this framework. We check the validity of this approximation in describing the non-stationary dynamical regime of several Ising models defined on Erdos-R\'enyi random graphs: we study ferromagnetic models with symmetric and partially asymmetric couplings, models with random fields and also spin glass models. A comparison with the actual Glauber dynamics, solved numerically, shows that one of the two studied approximations (the so-called `diamond' approximation) provides very accurate results in all the systems studied. Only for the spin glass models we find some small discrepancies in the very low temperature phase, probably due to the existence of a large number of metastable states. Given the simplicity of the equations to be solved, we believe the diamond approximation should be considered as the `minimal standard' in the description of the non-stationary regime of Ising-like models: any new method pretending to provide a better approximate description to the dynamics of Ising-like models should perform at least as good as the diamond approximation.
\end{abstract}

\maketitle

\section{Introduction}
\label{sec:Intro}
Dynamics is an important issue in almost every field of science, ranging from physics and biochemistry to neuroscience and social engineering \cite{barthelemy2008dynamical,castellano2009statistical,boccaletti2006complex}.
Nature and society have shown to be rich of systems presenting collective behaviour of many interacting agents. Neural networks and brain behaviour, gene regulatory networks, flocking or generally living systems and active matter are just few examples. Within statistical mechanics, a fundamental theory for the study of these systems, a satisfactory description of the time evolution of a many-particle system remains one of the most difficult subjects \cite{vankampen,balescu1975equilibrium}.
The core challenge is that even in cases where the microscopic processes guiding the dynamics are given, going from a very
general statement like a master equation to a practical solution is usually  unfeasible. This is due to the unavoidable difficulties of 
the exponential growth of the size of the state space with the number of particles and time intervals considered \cite{del2015dynamic,lokhov2014dynamic}.

For what concerns graphical models \cite{koller2009probabilistic,wainwright2008graphical}, as for instance disordered model defined on graph topologies \cite{mezard1990spin,mezard2009information}, in recent years there has been a sustained effort in the modelling of their dynamical behaviour for both dense and dilute networks \cite{semerjian2004stochastic}. Many concepts have been introduced in a natural analogy with the equilibrium theory, e.g. dynamical replica analysis \cite{hatchett2005dynamical,mozeika2009dynamical},  cavity method \cite{kiemes}, dynamic message-passing algorithm \cite{Neri2009,aurell2012dynamic,del2015dynamic,lokhov2014dynamic}, large deviation \cite{del2014perturbative,altarelli2013large},  TAP approaches \cite{roudi2011dynamical} and extended Plefka expansion for continuous variables \cite{bravi2015extended}.
Despite all these advances, the issue is far from being settled and there is an active community searching for approximate methods that accurately reproduce
numerical results from stochastic simulations \cite{barthel2015matrix}.

Recently Pelizzola in Ref.~\cite{pelizzola2013variational} has extended a simple variational technique based on a generalization of the cluster variational method (CVM) \cite{kikuchi61} to describe the stationary dynamical regime of Ising-like systems. The results obtained in Ref.~\cite{pelizzola2013variational} are extremely accurate, given the simplicity of the equations to be solved.

Here we claim that the same approximation based on CVM should be equally good in describing the non-stationary/transient regime, which is far more important in many applications, where the system under study is strongly out-of-equilibrium or in a changing environment.

To prove our claim we compare the analytical approximate solution with the actual Glauber dynamics solved running a large number of Monte Carlo (MC) simulations.
The comparison we make is extremely accurate since we check several microscopic observables (single spin magnetizations and two-point correlations in space and time), and not only macroscopic observables (global magnetization and energy).

The manuscript is organized as follows. In section \ref{sec:variational_kik_pel} we review the main ideas of a variational formulation of dynamics in
discrete time. In the following section \ref{sec:kinetic} we define the kinetic Ising model and its dynamic evolution.
In section \ref{sec:results} the main numerical results obtained after applying the variational formulation to this model are presented
and discussed in relation to MC simulations and another recently proposed approximation called dynamic message passing (DMP) with 1-step Markov memory \cite{del2015dynamic}. Finally, we discuss our findings in section \ref{sec:discussion} and close with two appendices including complementary technical details.

\section{A variational formulation of dynamics in discrete time}
\label{sec:variational_kik_pel}

In this section we briefly sketch  the approach formalized by R. Kikuchi in \cite{kikuchi61} and recently adapted and improved by A. Pelizzola in \cite{pelizzola2013variational}. 
The method relies on a generalization of the equilibrium cluster expansion
technique (also known as cluster variational method \cite{kikuchi,pelizzola05}) to include dynamical processes. One of the advantages of this approach is that it gives a scheme for
a hierarchy of approximations with increasing accuracy, always deducing the dynamical equations from a variational principle.
Hereafter we will follow the notation of \cite{pelizzola2013variational}.

Let $\mathbf{s}^t=\left\lbrace s_1^t\ldots s_N^t\right\rbrace$ be the set of variables that describe the state of a system
at time $ t\in \left[0\ldots t_f\right]$. If the evolution in time is stochastic, all the statistical information up to time $t_f$
is contained in the joint probability distribution of the histories, $P(\bs^0,\ldots,\bs^{t_f})$. This object is in itself untractable for large systems
since it takes a number of values $O(A^{N\times t_f})$ where $A$ is the typical cardinality of the variable $s_i$. Several probability distributions
depending on different subsets of variables will be used in this paper, all being marginals of the master probability $P$. 
In order to lighten notation, all of them will be written with the symbol $P$ and distinguished only by their arguments. 

We will focus on the commonly studied case of Markovian dynamics: 
\begin{equation}
P(\bs^0,\ldots,\bs^{t-1},\bs^t)=W(\bs^t|\bs^{t-1}) P(\bs^0,\ldots,\bs^{t-1})
\label{eqn:markov1}
\end{equation}
or, equivalently:
\begin{equation}
 P(\bs^t)=\sum_{\bs^{t-1}}W(\bs^t|\bs^{t-1}) P(\bs^{t-1}), \;\; \qquad\mbox{given} \:\:P(\bs^0). \label{eqn:markov2}
\end{equation}
Equations \eqref{eqn:markov1} and \eqref{eqn:markov2} are assumed valid for any $t\in [1..t_f]$.

In Physics it is always convenient to derive the fundamental relations from a variational principle. The cost in terms of abstraction
is greatly compensated by the comprehension of the internal structure of the theory in question. In this case, the central role is played by a 
functional $\mathcal{F}\left[P(\bs^0,\ldots,\bs^{t_f})\right]$ introduced by Kikuchi in \cite{kikuchi61} and defined as:
\begin{eqnarray}
  \mathcal{F}\left[P(\bs^0,\ldots,\bs^{t_f})\right]=\sum_{\bs^0,\ldots,\bs^{t_f}}P(\bs^0,\ldots,\bs^{t_f})
  \left[- \sum_{t=1}^{t_f} \ln W(\bs^{t}| \bs^{t-1})+\ln P(\bs^0,\ldots,\bs^{t_f})\right] 
  \label{eqn:kikuchifreeenergy}
\end{eqnarray}
The functional $\mathcal{F}$ takes as an argument the joint probability distribution $P$ and has a structure resembling that of a Gibbs free energy from equilibrium statistical mechanics.
The most interesting property of (\ref{eqn:kikuchifreeenergy}) is that if it is minimized in the space of $P$ taking into account the marginalization constraint:
\begin{equation}
\sum_{\bs^1,\ldots,\bs^{t_f}}P(\bs^0,\ldots,\bs^{t_f})=P(\bs^0)
\label{eqn:marg_constraint}
\end{equation}
the time evolution equation (\ref{eqn:markov1}) is recovered. This is, the probability distribution that minimizes $\mathcal{F}$ is
actually the one corresponding to the dynamic of the system.
For completeness this procedure is described in Appendix \ref{app:kikuchi_minimization}.     
     
The technical difficulty of computing $P(\bs^0,\ldots,\bs^{t_f})$ exactly is not reduced by the previous result, but it suggests a possible source of approximations. 
For example, some kind of mean field or factorization of $P$ can be proposed which amounts to minimizing $\mathcal{F}$ in a restricted subspace of distributions.     
In Kikuchi's original paper \cite{kikuchi61}, the joint probability distribution was parametrized in terms of two-times and single time probabilities. Cluster expansion
of the functional was then used to find approximations to the real probability distribution. An approach that appears more promising was proposed recently in 
\cite{pelizzola2013variational}. The latter, also inspired by the cluster variational method, makes an approximation to (\ref{eqn:kikuchifreeenergy}) obtained as a sum of 
the contributions of similar functionals written for the most correlated variables. Let us see this in more detail for a specific example.

In what follows we focus on locally tree-like topologies (i.e.\ random graphs) since we are interested in applications where dynamical discrete variables do interact via a diluted graphical model.
For our purposes, it is useful to think at the dynamic evolution as a set of copies of the original system, one for each time, that interact according to the transition matrix $W(\bs^t|\bs^{t-1})$. Furthermore, for a model with short range interactions and a Markovian dynamics the probability distribution of $s_i^t$ depends only on the previous time state of the variables it interacts with ($s_{\partial i}^{t-1} \equiv \{s_k^{t-1}\}_{k\in\partial i}$), where $\partial i$ denotes the subset of variables neighbours of $i$.
For ease of computation, we assume the state of $s_i^t$ not to depend on $s_i^{t-1}$: this is what happens, for example, in the heat-bath dynamics. Dependence of $s_i^t$ on $s_i^{t-1}$ can be introduced without any conceptual change \cite{PelizzolaPretti2017} and must be included if one is interested in other dynamics, as for example those in epidemic models.

We also assume that the spin transitions are independent events, or equivalently, that the transition matrix can be factorized as follows
\begin{equation}
  W(\bs^t|\bs^{t-1})=\prod_{i=1}^{N}  W_i(s_i^t|s_{\partial i}^{t-1}).
  \label{eq:transition_matrix_factorized}
\end{equation}
This choice corresponds to the so-called `parallel dynamics' in Monte Carlo simulations.
Under the above assumptions we can write
\begin{equation}
P(s_i^t,s_{\partial i}^{t-1})=W_i(s_i^t|s_{\partial i}^{t-1}) P(s_{\partial i}^{t-1})\,.
\label{eqn:exact_evolution}
\end{equation}

According to the prescription of the CVM, a first attempt to approximate the complete $\mathcal{F}$ may start from approximating the probability distribution in \eqref{eqn:kikuchifreeenergy} as a product of cluster probabilities. In the case of Markovian dynamics, we expect that the largest correlations between the variable $i$ at time $t$ and its neighbours at the previous time are encoded in clusters $A_i^{t}=(s_i^t,s_{\partial i}^{t-1})$. Therefore, following the CVM prescription, one can take $A_i^{t}$ as maximal cluster and expand the entropy term in \eqref{eqn:kikuchifreeenergy}. The result is a new definition for an approximated functional $ \mathcal{F}_S$ that is variational in the set of all cluster probability distributions $\left\lbrace P(s_i^t,s_{\partial i}^{t-1})\right\rbrace_{t=1,\ldots,t_f}$:
\begin{eqnarray}
 \mathcal{F}_S[\left\lbrace P(s_i^t,s_{\partial i}^{t-1})\right\rbrace_{t=1,\ldots,t_f}]=&&
 \sum_{i,t>0}^{t_f} \sum_{s^t_i,s_{\partial i}^{t-1}} P(s_i^t,s_{\partial i}^{t-1})
      \left[- \ln W_i(s^{t}_i| s^{t-1}_{\partial i})+\ln P(s_i^t,s_{\partial i}^{t-1})\right]\nonumber\\ 
 &&-\sum_{i,t>0}^{t_f-1} d_i \sum_{s^t_i} P(s_i^t)\ln P(s_i^t)\label{eqn:starfreeenergy}
 -\sum_{i} (d_i-1) \sum_{s^0} P(s_i^0)\ln P(s_i^0)\,,
\end{eqnarray}
where $d_i$ is the degree of vertex $i$, that is the number of neighbors of spin $s_i$.
First note that the first term on the RHS of (\ref{eqn:starfreeenergy}) is just a sum of functionals identical in structure to (\ref{eqn:kikuchifreeenergy}) but 
each one restricted to the set of variables $A_i^{t}=(s_i^t,s_{\partial i}^{t-1})$. These sets, in the CVM language, are the maximal regions
at this level of approximation. The meaning of the second and third term is simple: since
the sets $A_i^{t}$ overlap, some single variable contributions must be substracted in order to count each only once. This is the standard situation
in the context of CVM. This particular choice of variables included in $A_i^{t}$ is called \textit{star} approximation in \cite{pelizzola2013variational}.

The next step is to minimize this functional constrained to a set of consistency relations which are equivalent to (\ref{eqn:marg_constraint}):
\begin{eqnarray}
 P(s_i^t)&=& \sum_{s_{\partial i}^{t-1}} P(s_i^t,s_{\partial i}^{t-1})\\
 P(s_j^{t-1})&=& \sum_{s^{t}_i,s_{\partial i \setminus j}^{t-1}} P(s_i^t,s_{\partial i}^{t-1}).
\end{eqnarray}
The final result is that the minimizing probabilities obey the following equations:
\begin{eqnarray}
 P(s_i^t,s_{\partial i}^{t-1})&=& W_i(s^{t}_i| s^{t-1}_{\partial i})\prod_{j\in \partial i} P(s_j^{t-1})\label{eqn:star_evolution1}\\
 P(s_i^t)&=&\sum_{s_{\partial i}^{t-1}} W_i(s^{t}_i| s^{t-1}_{\partial i})\prod_{j\in \partial i} P(s_j^{t-1})
 \label{eqn:star_evolution2}
\end{eqnarray}
A convenient feature of equation (\ref{eqn:star_evolution2}) is that once the probabilities at a given time are known, the next generation of distributions
are generated using this simple prescription. The aforementioned relations have been obtained before \cite{hard_spin,Petermann,DeLosRiosPetermann} as an improved mean field theory 
but the approximations were then made on intuitive grounds. 
The comparison between the approximate equation \eqref{eqn:star_evolution1} and the exact dynamics \eqref{eqn:exact_evolution} shows that the star approximation amounts to assume that all the spins $s_{\partial i}$ neighbours of $i$ are probabilistically independent and therefore one expect it to be more accurate in the high temperature regime.

Much in the same way the \textit{diamond} approximation \cite{pelizzola2013variational} is derived. The fundamental idea is to include longer correlations in the time dynamics. This can be done, to some extent, taking into account correlations coming from $s_{\partial i}^{t-1}$ and previous interactions with other variables in the network. As can be easily seen, all variables in $s_{\partial i}^{t-1}$ 
interact with $s_{i}^{t-2}$ so, for the variable $i$, this should be the main source of correlation. The new region-based functional will take as fundamental elements all the groups in the maximal set, the diamond cluster  $B_i^{t}=\left\lbrace s_i^{t}\cup s_{\partial i}^{t-1}\cup s_i^{t-2}\right\rbrace $. So, following the CVM recipe, one can approximate the full joint probability by a product of cluster probabilities with the new maximal set $B_i^{t}$. The final result after constraint minimization reads

\begin{equation}
 P(s_i^t,s_{\partial i}^{t-1},s_i^{t-2})=W_i(s^{t}_i| s^{t-1}_{\partial i})\left[\prod_{j\in \partial i} P(s_j^{t-1},s_i^{t-2})\right] \left[P(s_i^{t-2})\right]^{1-d_i}
 \label{eqn:diamond_evolution1}
 \end{equation}
 and can be turned, alternatively, in
 \begin{equation}
 P(s_i^t,s_{\partial i}^{t-1},s_i^{t-2})=W_i(s^{t}_i| s^{t-1}_{\partial i})\left[\prod_{j\in \partial i} P(s_j^{t-1}|s_i^{t-2})\right] P(s_i^{t-2}),
 \label{eqn:diamond_evolution2}
\end{equation}
where the standard definition for conditional probabilities  $P(s_j^{t-1},s_i^{t-2}) = P(s_j^{t-1}|s_i^{t-2})  P(s_i^{t-2})$ is used.

As for the previous cluster expansion, we get a set of equations that can be iterated in time. The results obtained using this second proposal are expected to improve those from the star approximation, the reason
being that the factorization in the star anzats is refined by conditioning on the state of the common neighbor, compare (\ref{eqn:star_evolution1}) and (\ref{eqn:diamond_evolution2}). 

Let us observe that neither the maximal cluster in the star approximation, i.e. $A_i^{t}=(s_i^t,s_{\partial i}^{t-1})$, nor the maximal cluster in the diamond approach $B_i^{t}=\left\lbrace s_i^{t}\cup s_{\partial i}^{t-1}\cup s_i^{t-2}\right\rbrace $ includes the state $s_i^{t-1}$. This is because one of our working assumptions is that the state of a spin at time $t$ does not depend on the same spin at time $t-1$. For other dynamical rules where the state $s_i^{t-1}$ can determine directly the state of $s_i^{t}$, e.g. in epidemic models, one would need to include also $s_i^{t-1}$ in the cluster constructions to account the effect of such interaction \cite{PelizzolaPretti2017}.

\section{The kinetic Ising model}
\label{sec:kinetic}
The numerical test of the approximations introduced in the previous section that we perform, in this contribution, is made for the kinetic Ising model, typically used as a prototype to investigate spin dynamics.
This model is defined as a set of $N$ Ising spins $s_i=\pm1$ placed on the vertices of a graph $G$ that
describes the topology of the interactions $J_{ij}$, plus a rule for the time evolution of these variables \footnote{In general $G$ could be a directed
graph and $J_{ij}\neq J_{ji}$}. We will consider a parallel Glauber dynamic \cite{Glauber63} by means of a transition matrix 
of the form \eqref{eq:transition_matrix_factorized}, where for each spin $i$ we have:
\begin{equation}\label{eq:rate_w}
W_i(s_i^t|s_{\partial i}^{t-1})=
\dfrac{\exp \left[\beta s_i^t\left(h_i^t +  \sum_{j\in \partial i}  J_{ji} s_j^{t-1}\right) \right]}
{2\cosh \left[\beta \left(h_i^t +  \sum_{j\in \partial i}  J_{ji} s_j^{t-1}) \right)\right]}
\end{equation}
with $\beta$ and $h_i^t$ being respectively the inverse temperature and an external local field at time $t$. The behavior of this model depends essentially on two features. On the one hand, the topology of the interaction graph $G$, whether it is a lattice, 
a random graph, fully connected, etc, and on the other, the symmetry of the interactions $J_{ij}$. Hereafter, according to the literature, we denote \emph{symmetric} those graphs having $J_{ij} = J_{ji}$ and \emph{partially asymmetric} or \emph{asymmetric} those graphs for which $J_{ij} \neq J_{ji}$. Depending on these properties, the system
may, for example, not satisfy detailed balance conditions or reach a stationary state different from thermal equilibrium \cite{aurell2011message}. In any case, 
the variational formalism can be straighforwardly applied to different levels of approximation. In the rest of this section we present the results of the star and the diamond
for  this model.

All equations in the star approximation can be expressed in terms of single site probabilities, which are parametrized using local magnetizations: $P(s_i^t)=\frac{1+m_i^t s_i^t}{2}\;$ where $\;m_i^t=\sum_{s_i^t} s_i^t P(s_i^t)$.
Local magnetizations are then all the information  that is kept at this level. The time propagation in this case can be recast from 
(\ref{eqn:star_evolution2}) in the following compact form:
\begin{equation}
m_i^t=\sum_{s_{\partial i}^{t-1}}\tanh \beta \Big( h_i^t +  \sum_{j\in \partial i}  J_{ji} s_j^{t-1} \Big) \prod_{j\in \partial i} \dfrac{1+m_j^{t-1}s_j^{t-1}}{2}.
\end{equation}
Some non trivial correlations can be estimated once we obtain single site magnetizations, for example, nearest neighbour disconnected correlation at consecutive times is directly 
derived from equation (\ref{eqn:star_evolution1}) if $k\in \partial i$
\begin{equation}\label{eq:star_corr}
 c_{i,k}^{t,t-1}=\langle s_i^{t}s_k^{t-1} \rangle=\sum_{s_{\partial i}^{t-1},s_i^{t}}s_i^{t}s_k^{t-1}\W{s_i^t}{s_{\partial i}^{t-1}}
				    \prod_{j\in \partial i}\dfrac{1+m_j^{t-1}s_j^{t-1}}{2}
\end{equation}
and the connected correlation can be computed accordingly to
\begin{equation}
 \left(c_{i,k}^{t,t-1}\right)_c=c_{i,k}^{t,t-1}-m_i^t m_k^{t-1}
\end{equation}
For the diamond approximation, in addition to single site probabilities we need to consider the joint distribution of nearest neighbors
at consecutive times. The evolution of the system is reduced to the propagation in time of a set of equations relating these variables (see (\ref{eqn:diamond_evolution1})):
\begin{equation}
P(s_i^t,s_j^{t-1})=\sum_{s_{\partial i \setminus j}^{t-1},s_i^{t-2}}W_i(s^{t}_i| s^{t-1}_{\partial i})\left[\prod_{j\in \partial i} P(s_j^{t-1},s_i^{t-2})\right] \left[P(s_i^{t-2})\right]^{1-d_i}
 \label{eqn:diamond_evolution_marginalized}
\end{equation}
The correlation between nearest neighbors is, in this case, part of the set of variables and not a deduced quantity as in the star case. This
is a fundamental advantage that will provide much more accurate predictions, as will be shown with numerical simulations in the Section \ref{sec:results} . An algorithm can be easily
written to iterate equation (\ref{eqn:diamond_evolution_marginalized}). Alternatively, one can use the magnetization and correlation instead of
propagating probabilities, which are always a bit redundant because of normalization constraints. The choice is a tradeoff between space in memory (larger when 
storing probabilities) and simulation time (longer when marginalizing to find magnetization and correlation).
Probabilities and moments are related as usual by the following
\begin{equation}
 P(s_i^t,s_j^{t-1}) = \dfrac{1}{4}\left( 1 + m_i^t s_i^t + m_j^{t-1} s_j^{t-1} + c_{i,j}^{t,t-1}  s_i^t s_j^{t-1}\right)
\end{equation}
This last expression can be plugged in \eqref{eqn:diamond_evolution_marginalized} to obtain iterative equations for the magnetization and correlations which we use for the numerical implementation.

\section{Results}
\label{sec:results}
In this section we numerically test the accuracy of the star and diamond approximation illustrated in Section \ref{sec:variational_kik_pel} on the kinetic Ising model presented in Section \ref{sec:kinetic} and relate these results with MC simulations. The numerical analysis is done on a Erdos-R\'enyi random graph topology for two ferromagnetic models and two different disorder models as the Random Field Ising Model \cite{imry1975random} and the Viana-Bray Spin Glass \cite{viana1985phase}.
  
In general, the only exact procedure available for a statistical description of the set of states generated by the dynamic evolution of a system  
is precisely the explicit construction of this set of states. This is usually done via stochastic simulations, i.e.\ Monte Carlo (MC), or molecular dynamics.
Accuracy in both methods is obtained by paying a cost in terms of computational effort. For example, in non-equilibrium MC simulations, accurate averages are
computed by summing over a very large number of dynamical trajectories. This amounts to run the algorithms many times with different choices for the
initial conditions and different sets of random numbers for the acceptance-rejection rule. All this has a cost that increases linearly with the
number of runs $N_r$, whereas statistical errors decrease as $N_r^{-1/2}$. 

At this point approximate algorithms like the ones described in the previous sections become very useful. These are one-run algorithms in 
the sense that, given the initial condition $P(s^0)$, the corresponding set of equations is solved only once, moving forward in time. This simplification
comes at the cost of having only a reduced set of parameters, like the local magnetization or nearest neighbours correlations, to describe
the statistics of the systems. Nevertheless, it is worth investigating the conditions and models where this approximations can be useful.
In \cite{pelizzola2013variational} some results are shown for the stationary behaviour of the star and diamond approximation in contrast to MC
simulations and other equivalent methods. A kinetic Ising spin model with asymmetric interactions is analyzed on finite dimensional lattices as well
as on a random regular graph. Those results show that for the stationary state (for asymmetric interactions equilibrium in the thermodynamical 
sense is not attained) the diamond approximation gives the best estimates of single site magnetization among all approaches considered. In what follows we show that this family of approximations based on the CVM well describes also the transient regime of the microscopic variables for 
some symmetric and asymmetric tree-like networks.

\subsection{Symmetric and partially asymmetric ferromagnet}\label{sec:ferromagnet}

In this subsection we compare numerical results for the star, diamond and DMP 1-step Markov memory \cite{del2015dynamic} approximations on an Erdos-R\'enyi random graph (ERRG) with $N=10^3$ sites and a mean connectivity $c=3$. The accuracy of these methods is then compared with MC simulations averaged on $10^6$ runs. 

The typical computational time of the approximations described in the previous sections with the aforementioned  simulation parameters is of order of a few minutes on a desktop PC whereas the MC simulation time takes much longer, of the order of several hours.

\begin{figure}[h!]
 \centering
 \subfloat[Global magnetization.]{
 \includegraphics[width=0.45\textwidth,keepaspectratio=true]{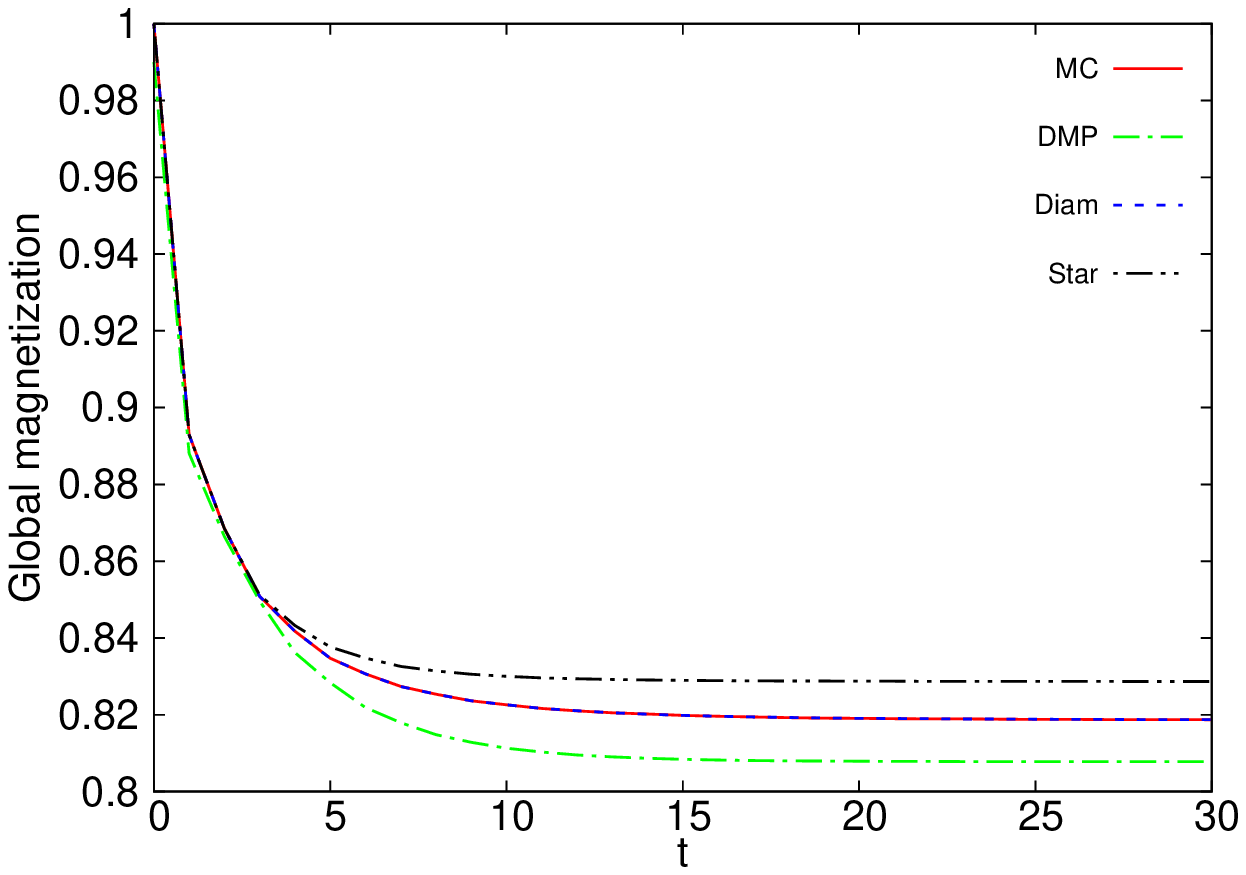}
 \label{fig:global_mag_sym_low}
 }
 \hspace{0.25cm}
 \subfloat[Distance from MC prediction.]{
\includegraphics[width=0.45\textwidth,keepaspectratio=true]{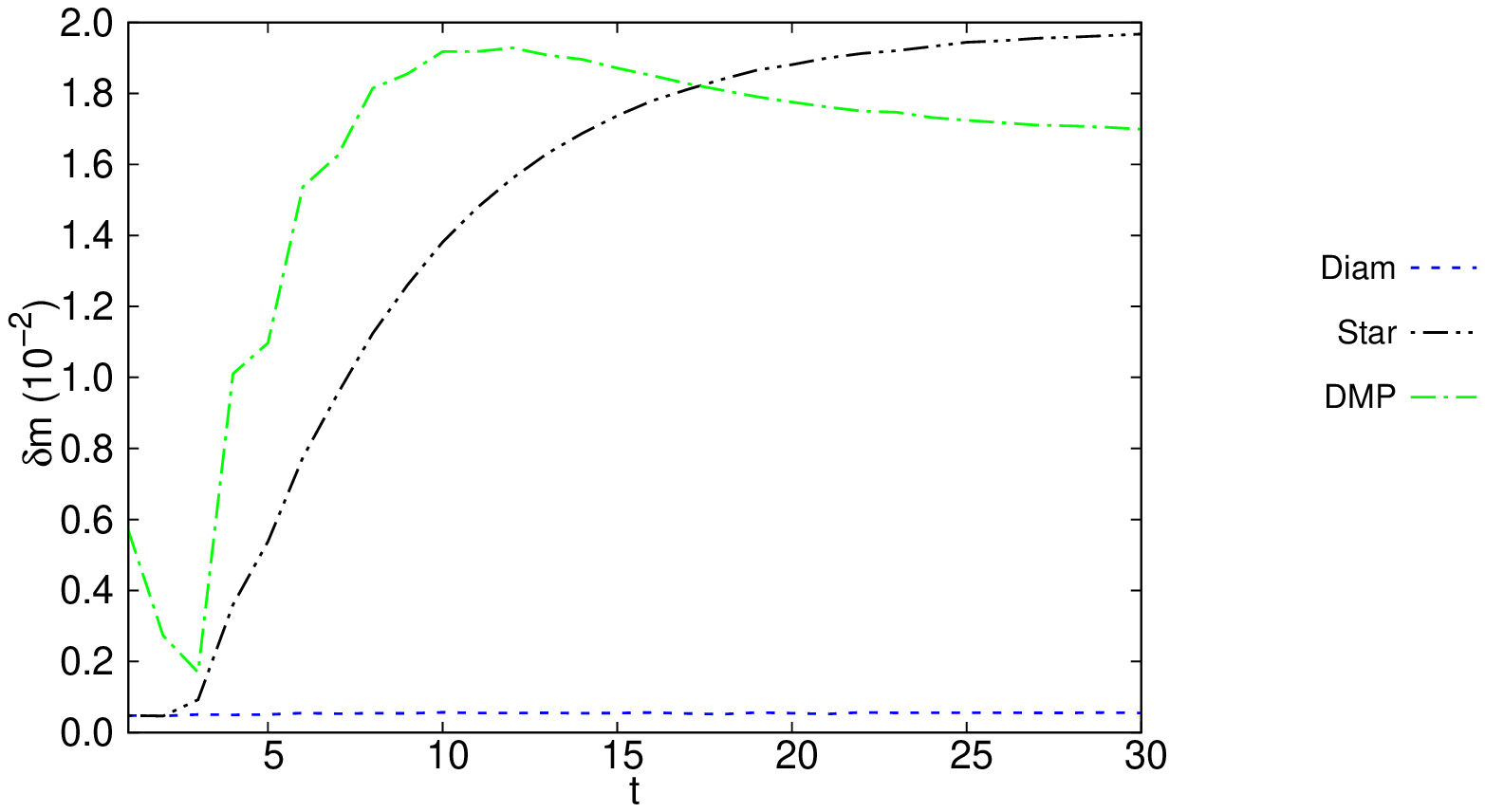}
 \label{fig:local_mag_error_sym_low}
 }
 \caption{Left panel: results for global magnetization on a network with symmetric interactions at temperature $T=0.5$ below the critical ferromagnetic transition $T_c = 0.962$ \cite{Leone2002}. Different lines refer to different methods. MC stands for Monte Carlo simulations averaged over $10^6$ runs, DMP for dynamic message-passing 1-step Markov memory, diamond and star for the approximations presented in the main text. At the initial time the configuration of the spins is such that the global magnetization $m_0=1$. Right
 panel: error in the estimation of local magnetizations in comparison with MC simulations (see \eqref{eqn:delta_m}).}
 \label{fig:ferro_magn}
\end{figure}

Simulations are started from a state far from equilibrium; all spins set in the same direction.
In Figure \ref{fig:global_mag_sym_low} we report the time evolution of the global magnetization for a ferromagnet with symmetric interactions ($J_{ij}=J_{ji}=\frac{1}{c}$) at low temperature. 
The predictions for the first time steps are almost the same for all methods considered. However, for longer times the diamond approximation
obtains a much better estimate of this global average. In the high temperature phase all the three approximations are almost indistinguishable to MC and therefore we do not report these results here.

%
%

%
%
\begin{figure}[t!]
\centering
\subfloat[Nearest neighbors correlation.]{
 \includegraphics[width=0.4\textwidth,keepaspectratio=true]{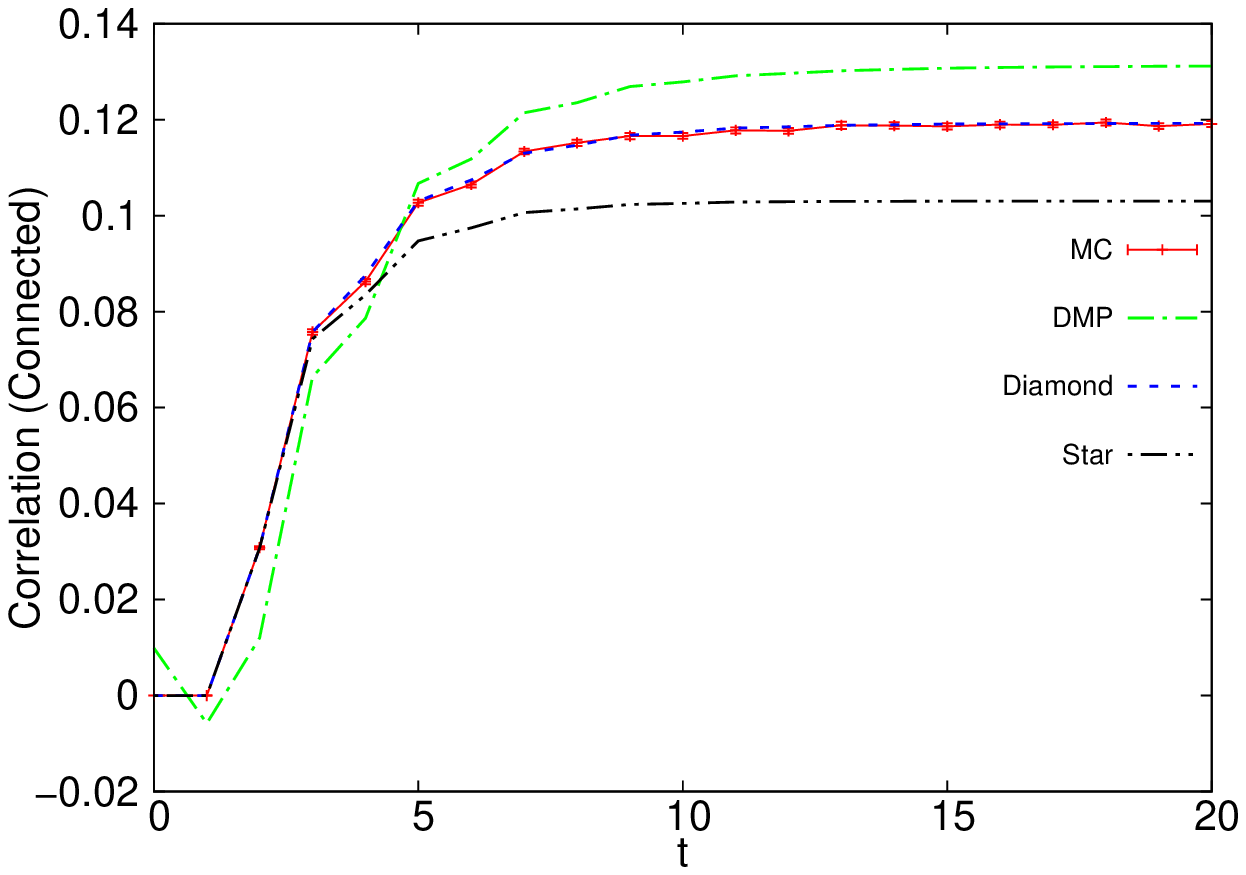}
\label{fig:corr_sym_low}}
\hspace{0.5cm}
 \subfloat[Two time steps connected autocorrelation.]{
 \includegraphics[width=0.4\textwidth,keepaspectratio=true]{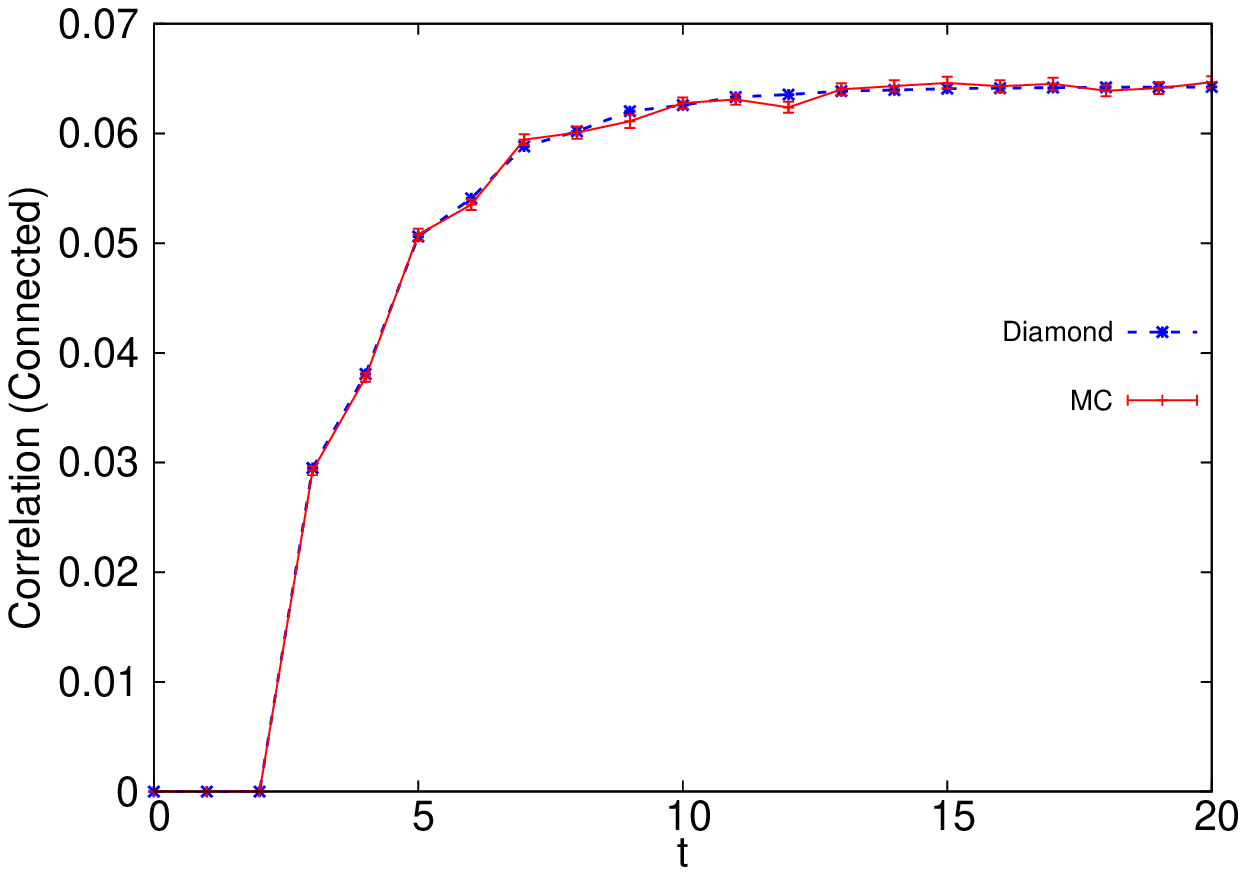}
\label{fig:autocorr_sym_low}}
 \caption{Symmetric network. Left panel: estimation of connected correlation function $C_{ij}(t,t-1)$ between a spin $i$ with degree equal to the mean degree $c=3$ at time $t$ and one of its neighbours $j$ at time $t-1$. Results refer to a given pair of spins.  Different lines refer to the different methods according to the caption of Figure \ref{fig:ferro_magn}. Right panel: autocorrelation of spin $i$ at two time steps, $C_i(t,t-2)$, estimated by the diamond approximation. Results refer to a given spin. In both panels the initial conditions are the same of Figure \ref{fig:ferro_magn}. Error bars correspond to the standard error of the mean estimated from $10^6$ MC runs.}
 \label{fig:corr_symm}
\end{figure}

To test the accuracy of the approximations for local observables, in Figure \ref{fig:local_mag_error_sym_low} we report a plot of an overall
measure of the distance between the set of approximated local magnetizations computed with the different approaches and the MC results for the same quantities. The initial conditions are the same as those in Figure \ref{fig:global_mag_sym_low}. 
The mean deviation $\delta m(t)$ is defined as:
\begin{equation}
 \delta m(t)=\sqrt {\frac{\sum_{i=1}^{N}(m_i^A(t)-m_i^{MC}(t))^2}{N}}
 \label{eqn:delta_m}
\end{equation}
where $m_i^A(t)$ stands for the approximated local magnetization by using one of the approaches reported. It confirms that the diamond approximation typically finds local magnetization with an error around $10^{-3}$ for a quantity that is $\mathcal{O}(1)$ in a ferromagnetic state. For a MC simulation with $10^6$ runs the statistical error for the local magnetization is also near $10^{-3}$ which means that the error made by the diamond approach is almost indistinguishable from MC fluctuations. The long term agreement for the
symmetric network is indeed not surprising since the diamond solution in the stationary case coincides with the belief propagation solution \cite{pelizzola2013variational}, known
to be very accurate on tree-like topologies. The transient behaviour, on the other hand, is usually more difficult to reproduce and this simple method gives very good estimates in this region.

Since the basic variable sets defined in the context of each approximation contain several spins at different positions and times, some two-point correlations
are easily derived. For instance, in Figure \ref{fig:corr_sym_low} nearest neighbour correlations are computed according to \eqref{eq:star_corr} for the star, to \eqref{eqn:diamond_evolution_marginalized} for the diamond and by using the formulation described in Appendix \ref{app:DMP} for the dynamic-message passing approach.
At this stage it is worth remembering that parallel dynamics runs two histories that are independent from each other, e.g. nearest neighbors \textit{at the same time} 
never have a common spin in their respective histories. Therefore the only possible source of correlation between them is the initial condition.
On the other hand, a spin and its nearest neighbour at one-time distance are strongly correlated since they appear simultaneously in the update
equations for the probabilities (see equation \eqref{eqn:exact_evolution}). Figure \ref{fig:corr_sym_low} shows that the diamond approximation, contrarily to the other methods, in addition to the magnetization well reproduces also the out-of-equilibrium behaviour of correlation functions for this symmetric model.
The same method also allows a straightforward computation of the autocorrelations at time $t$ and $t-2$, as reported in equation \eqref{eqn:diamond_evolution1}, and results for this quantity are shown in Figure \ref{fig:autocorr_sym_low}. The same observables cannot be computed by using the star approximation, because it is not included in any CVM region.
%

\begin{figure}[t!]
 \centering
 \subfloat[Global magnetization.]{
 \includegraphics[width=0.4\textwidth,keepaspectratio=true]{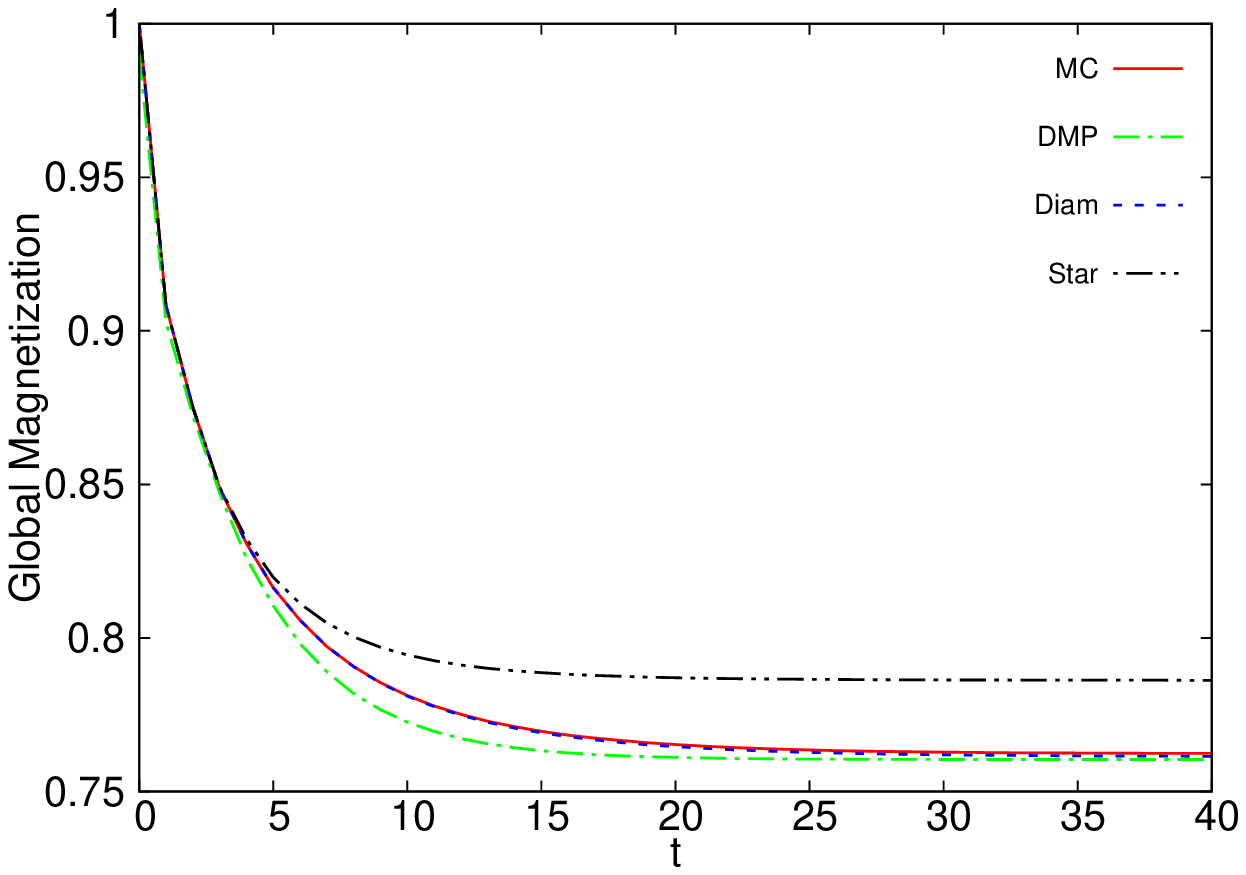}
 \label{fig:global_mag_asym_low}
 }
 \hspace{0.5cm}
 \subfloat[Distance from MC prediction.]{
 \includegraphics[width=0.4\textwidth,keepaspectratio=true]{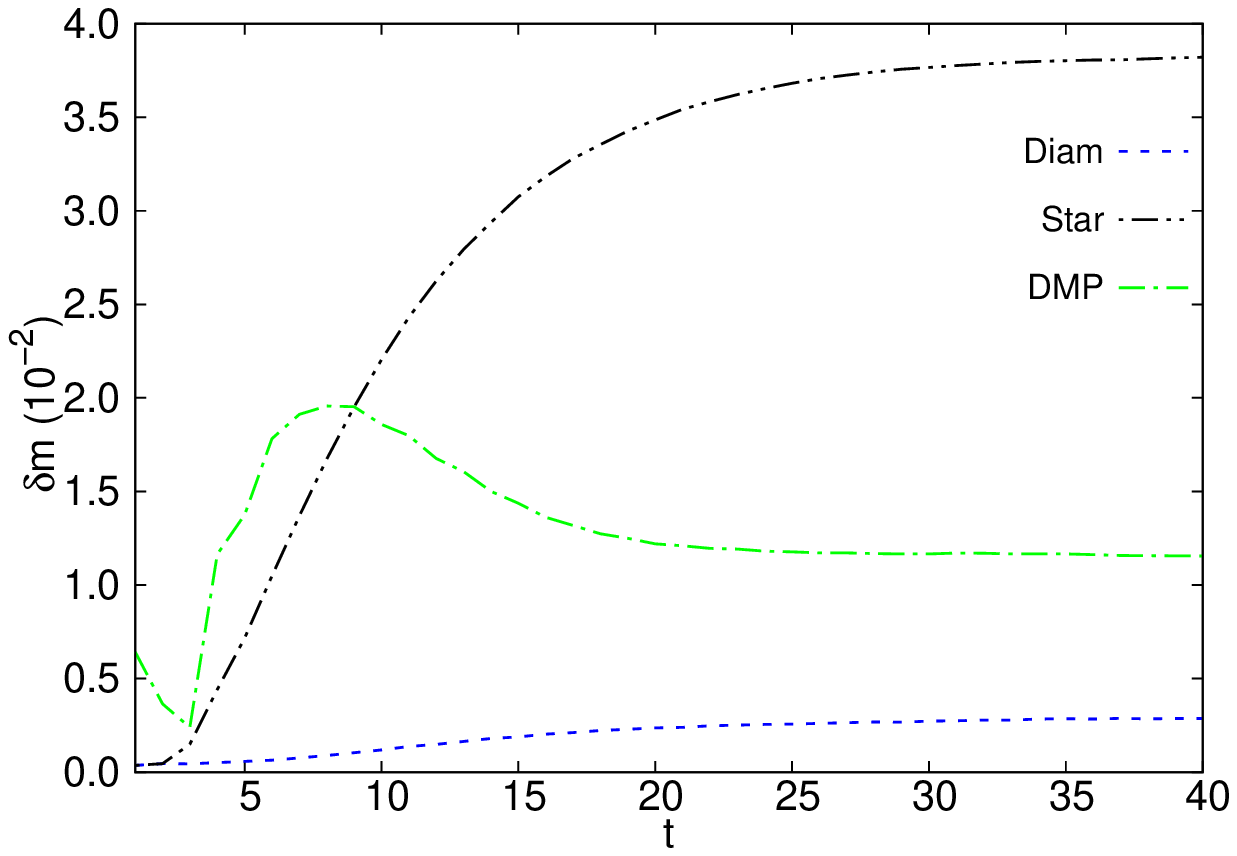}

 \label{fig:local_mag_error_asym_low}
 }
 \label{fig:asym_mag_error}
 \caption{Results for global and local magnetization of a kinetic Ising model with asymmetric interactions ($J_{ij}/J_{ji}=1/4$). The temperature, $T=0.833$, corresponds to a magnetized phase (in this model ergodicity breaks at $T_c \approx 1.7$). Left panel: dynamic evolution of the global magnetization. Different lines corresponds to the different approaches listed in Figure \ref{fig:ferro_magn}. Right panel: error in the estimation of local magnetizations in comparison with MC simulations (see \eqref{eqn:delta_m}).}
\end{figure}

A similar numerical analysis as the one presented for the ferromagnet with symmetric interactions can be done for asymmetric networks. We consider here the dynamic evolution of the kinetic Ising model for a partially asymmetric ferromagnet. This model is best described by a directed graph where interacting spins $(i,j)$ are connected by two directed edges with opposite directions and different interaction strengths, say $J_{ij}=1/c$ and $J_{ji}=1/4c$. For each edge, the direction with the stronger coupling is chosen uniformly randomly and independently from the other edges. Results for the dynamic evolution of the global magnetization at low temperature are illustrated in Figure \ref{fig:global_mag_asym_low}. The diamond approximation, as in the previous symmetric case, outperforms the other methods. 
Comparing to the symmetric ferromagnet case (see Fig.~\ref{fig:global_mag_sym_low}), the DMP approach with 1-step Markov memory improves the dynamic reconstruction, whereas the star approximation worsens. 
In the high temperature regime the star approximation quantitatively deviates from the MC simulations whereas both diamond and DMP provide more accurate results. We do not report these outcomes here because are of less interest compared to the low temperature case. In Figure \ref{fig:local_mag_error_asym_low} we also report the error in the computation of local magnetizations. As for the symmetric case, the diamond approximation outperforms the other approximations and typically finds local magnetization with a relative error of less than $0.5\%$. 

%
%

%
%
\begin{figure}[!t]
\centering
\subfloat[$C_{ij}(t,t-1)$ at high temperature]{
 \includegraphics[width=0.4\textwidth,keepaspectratio=true]{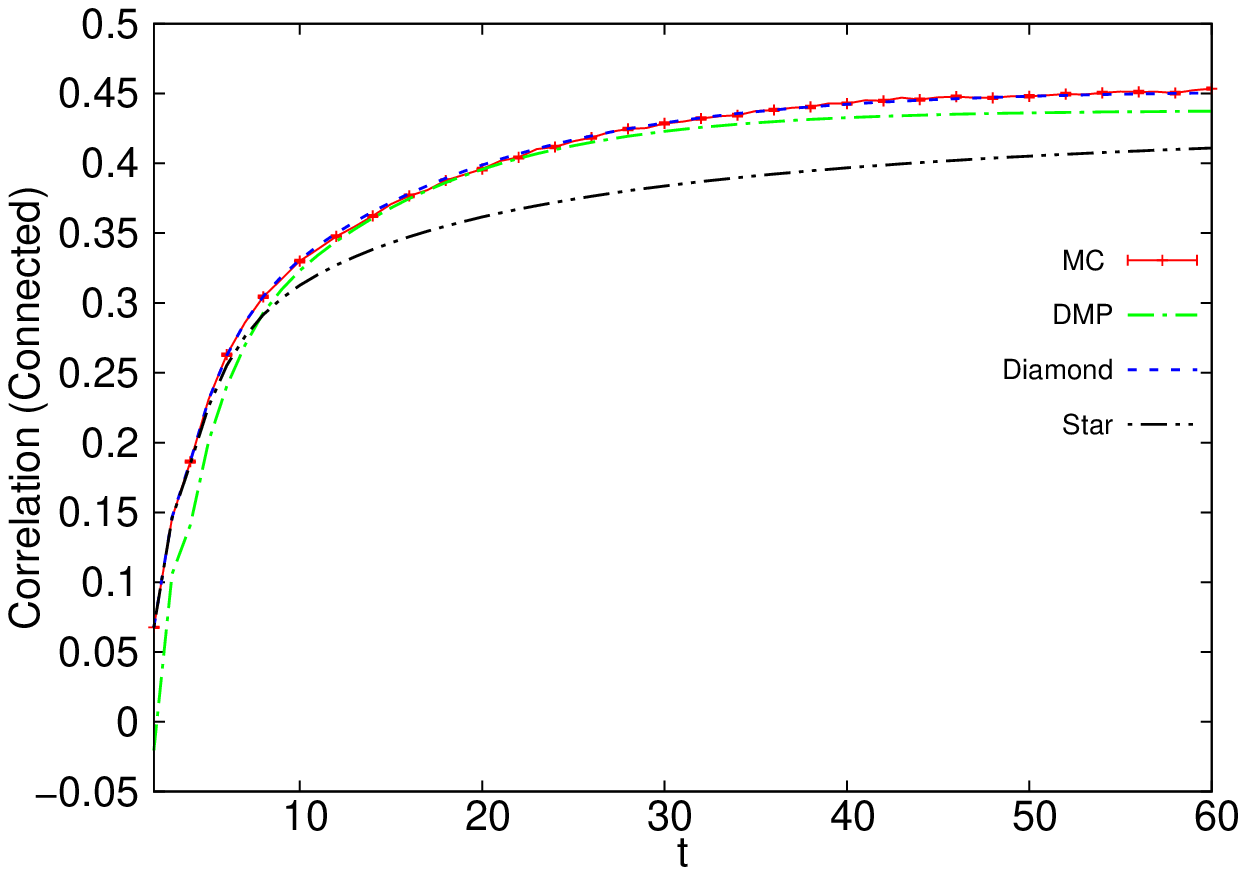}
\label{fig:corr_asym_high}}
\hspace{0.5cm}
 \subfloat[$C_{ij}(t,t-1)$ at low temperature]{
 \includegraphics[width=0.4\textwidth,keepaspectratio=true]{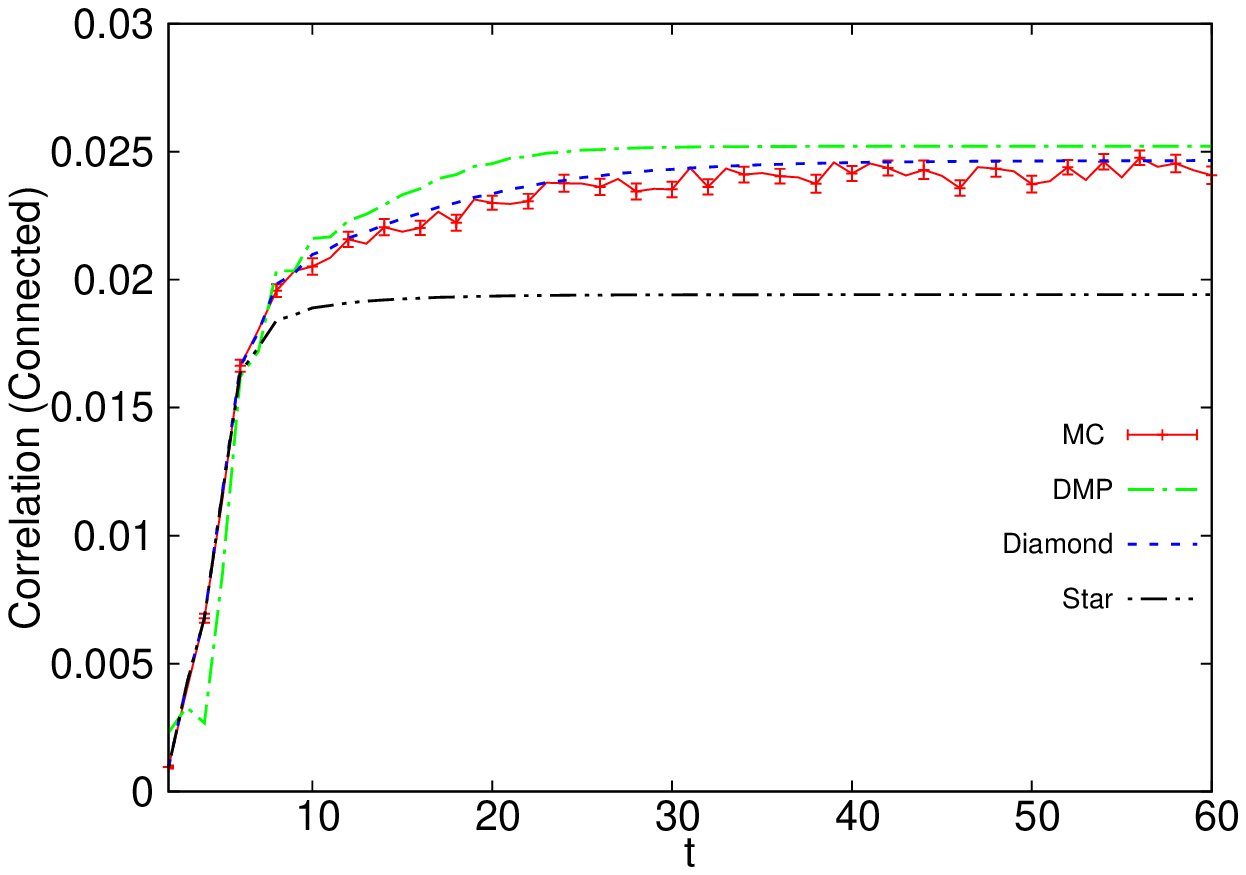}
\label{fig:autocorr_asym_low}}
 \caption{Partially asymmetric network ($J_{ij}/J_{ji}=1/4$). Estimation of the dynamical evolution of the connected correlation function $C_{ij}(t,t-1)$ between a given pair of spins, by using the different approaches discussed in the main text. Different lines refer to the different methods listed in Figure \ref{fig:ferro_magn}. Left panel: high temperature regime, $T=2$. Right panel: low temperature regime, $T=0.833$. Error bars correspond to the standard error of the mean estimated from $10^6$ MC runs.}
\end{figure}
The dynamics of connected correlations at high and low temperature for this model is reported respectively in Figure \ref{fig:corr_asym_high} and \ref{fig:autocorr_asym_low}. The star approximation deviates from MC simulations already for high temperatures and its numerical results gets progressively worsen by lowering the temperature. The DMP 1-step Markov memory performs better at high temperatures but deviate from MC simulations at low temperature, especially in the out-of-equilibrium transient. The diamond approximation performs very well in both cases being almost indistinguishable from MC simulations for the high temperature case. Note that the equilibration value of the correlation in Figure \ref{fig:corr_asym_high}
is not negligible, signaling the vicinity of a phase transition where correlations are larger.

\subsection{Random Field Ising Model and Viana-Bray Spin Glass}\label{sec:RFIM_SG}

In this subsection we report the numerical results obtained by applying the variational approach illustrated in Section \ref{sec:variational_kik_pel} on the Random Field Ising Model \cite{imry1975random} and the Viana-Bray Spin Glass \cite{viana1985phase} compared with the dynamic message-passing approach and MC simulations.

The RFIM is a paradigmatic disordered model where the disorder is not encoded topologically in the randomness of the coupling interactions $J_{ij}$ but rather in a random external local field $h_i$ acting on each spin within the network. It represents one of the simplest models that exhibits cooperative behaviour with quenched disorder and can be considered, somehow, complementary to the Ising spin glass. The energy function at equilibrium for this model is $H(\boldsymbol{s}) =- \sum_{(ij)} J s_i s_j - \sum_i h_i s_i$. 
The presence of the random external local field antagonizes the ordering effect due to ferromagnetic couplings and therefore one expects a lowering of the transition temperature increasing the magnitude of the local field. For low enough fields, or low enough temperatures, the system is found in a ferromagnetic phase, whereas, in the opposite limits, it is found in a paramagnetic one. For the dynamical simulations of this model, we use the Glauber transition rate of equation \eqref{eq:rate_w} with $J_{ij} = J_{ji} = 1/c$ and a random local external field constant in time, \emph{i.e.} $h_i^t = h_i = \pm 0.3$, extracted from a bimodal distribution. 

\begin{figure}[t!]
\centering
\subfloat[Global magnetization.]{
 \includegraphics[width=0.4\textwidth,keepaspectratio=true]{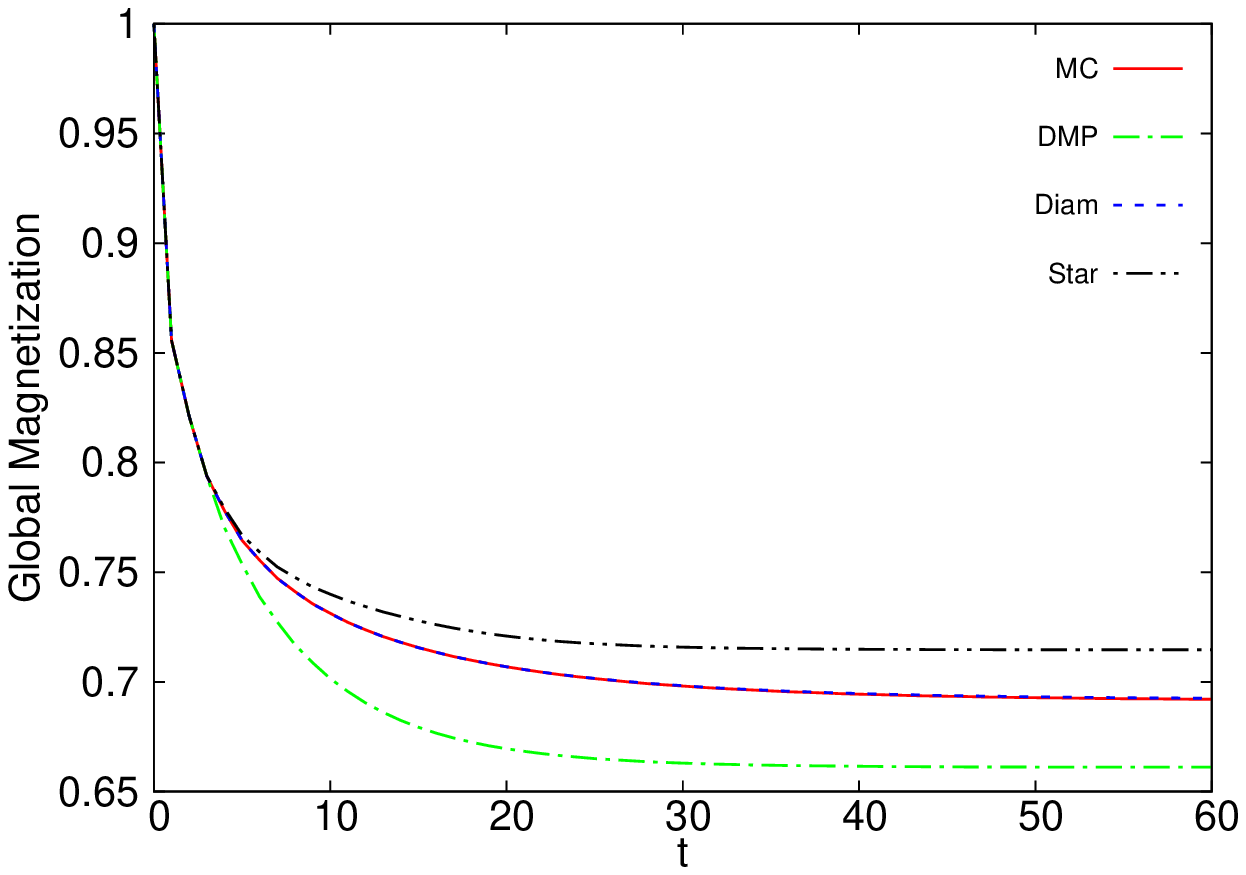}
\label{fig:magn_RFIM_low}}
\hspace{0.5cm}
 \subfloat[Nearest neighbours correlation.]{
 \includegraphics[width=0.4\textwidth,keepaspectratio=true]{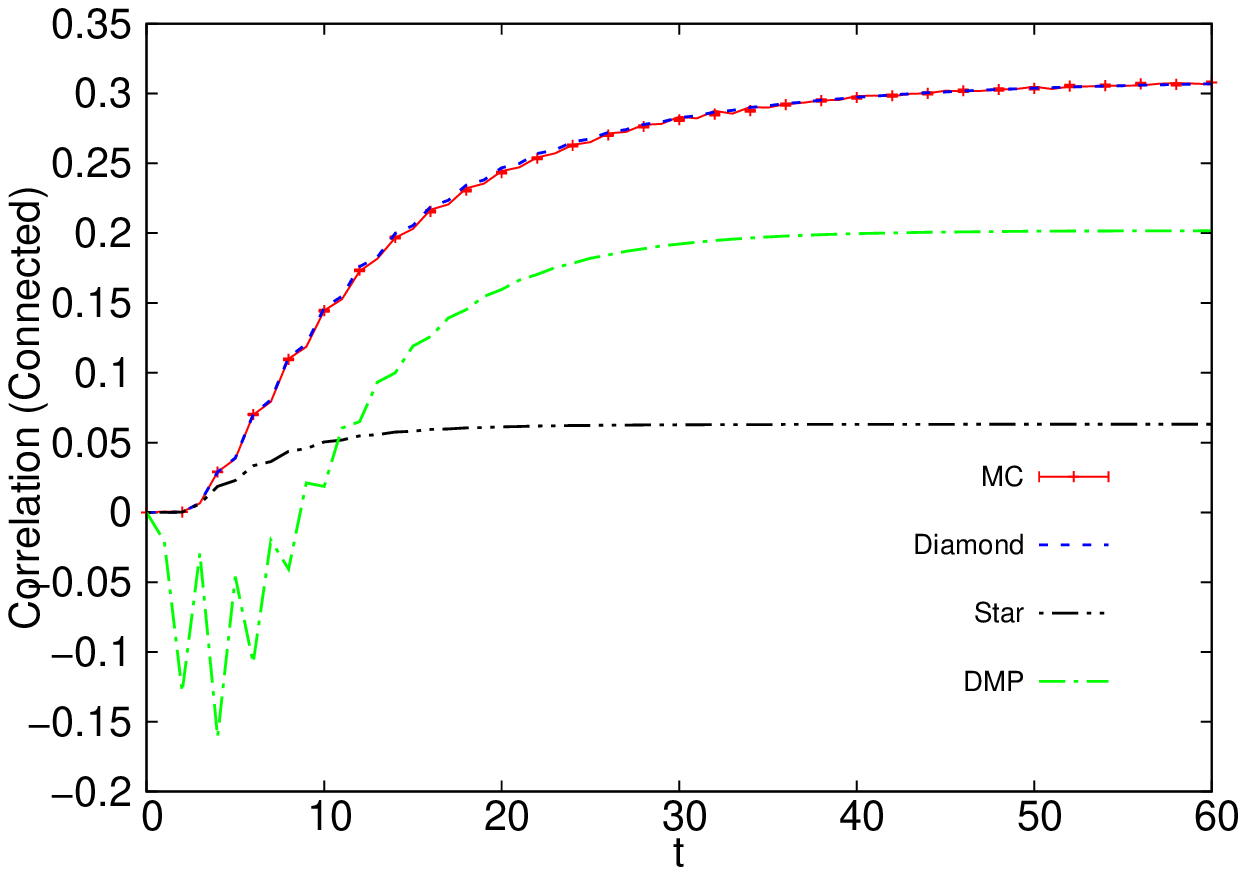}\label{fig:corr_RFIM_low}}
 \caption{Estimation of the dynamical evolution of the RFIM at temperature $T=0.5$ below the critical transition temperature $T_c(h=0.3)\approx 0.78$. Left panel: dynamics of the global magnetization. Different lines represent different methods presented in the main text and listed in Figure \ref{fig:ferro_magn}. Right panel: estimation of the dynamics for  the two-times connected correlation function between a randomly chosen spin $i$ and one of its neighbour $j$, \emph{i.e.} $C_{ij}(t,t-1)$.}
\end{figure}
Similarly to the symmetric ferromagnetic case of Section \ref{sec:ferromagnet}, the dynamics of the global magnetization at high temperature is well recovered by all the approximations at any time and therefore is not reported here.
In Figure \ref{fig:magn_RFIM_low} we show the global magnetization at low temperatures, \emph{i.e.} below the critical transition, which by a
population dynamics calculation \cite{Doria201558} can be estimated around $T_c=0.78$ for the value of $h_i$ used. Except for very short times, both the star and the DMP 1-step Markov memory approximation deviates from MC simulations whereas the diamond approximations well reproduces the behaviour of this observable at every time in the dynamics. Two-times connected correlation functions are illustrated in Figure \ref{fig:corr_RFIM_low}. The diamond approximation, also in this case, remains the most performing approximation, almost indistinguishable from MC simulations.

As last example for the numerics, we test the accuracy of the approximation on the dynamics of the Ising spin glass Viana-Bray model \cite{viana1985phase}. Contrarily to the RFIM, the Ising spin glass presents topological disorder in the \emph{quenched} couplings $J_{ij}$ which are        sampled randomly from either a gaussian or a bimodal distribution. The presence of both positive and negative couplings in the networks generates a very irregular energy landscape. This difference - respect to the previous analyzed models - enriches very much the physics of this system which shows a spin glass phase transition in addition to the ferromagnetic transition seen for the previous models.

\begin{figure}[t!]
\centering
\subfloat[Global magnetization at T= 0.5.]{
 \includegraphics[width=0.4\textwidth,keepaspectratio=true]{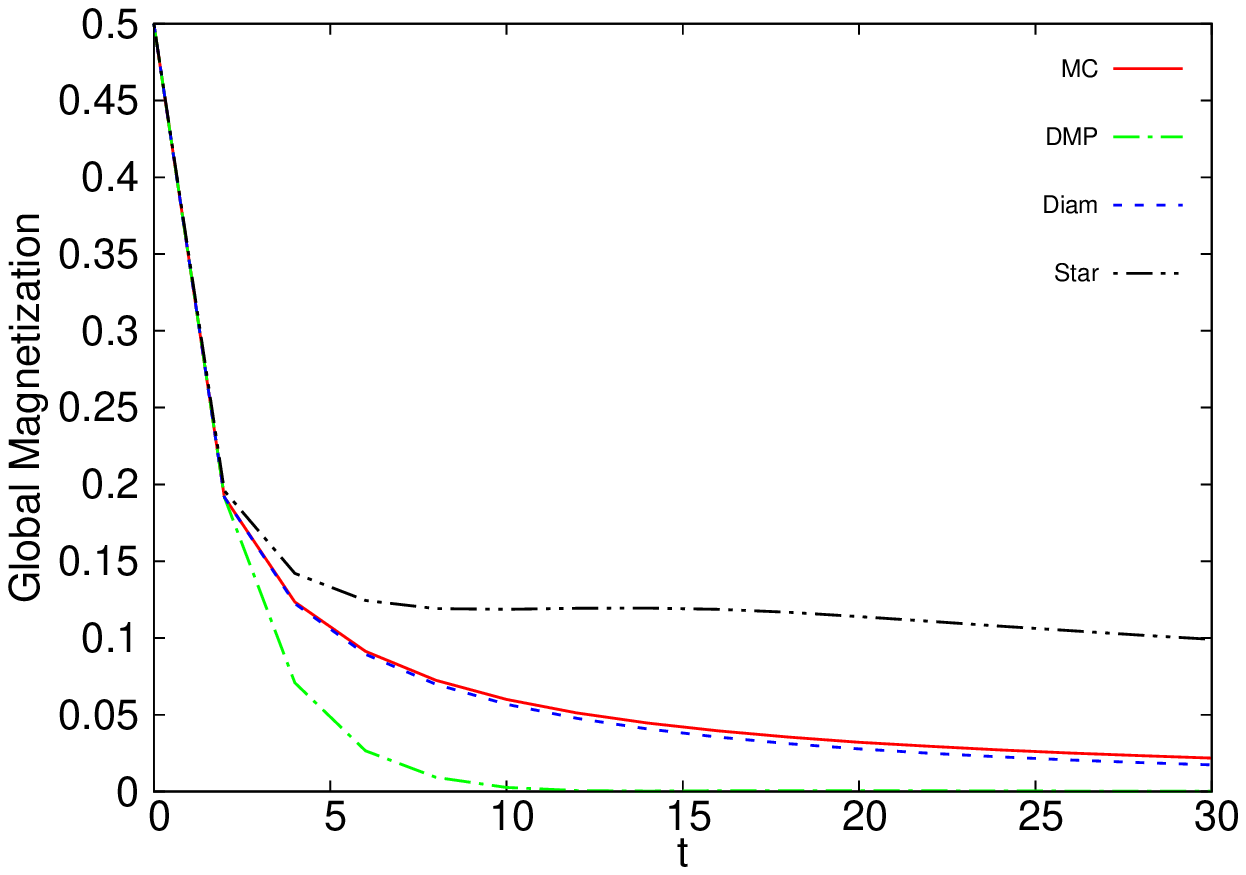}
\label{fig:magn_SG_high}}
\hspace{0.5cm}
 \subfloat[Global magnetization at T=0.25.]{
 \includegraphics[width=0.43\textwidth,keepaspectratio=true]{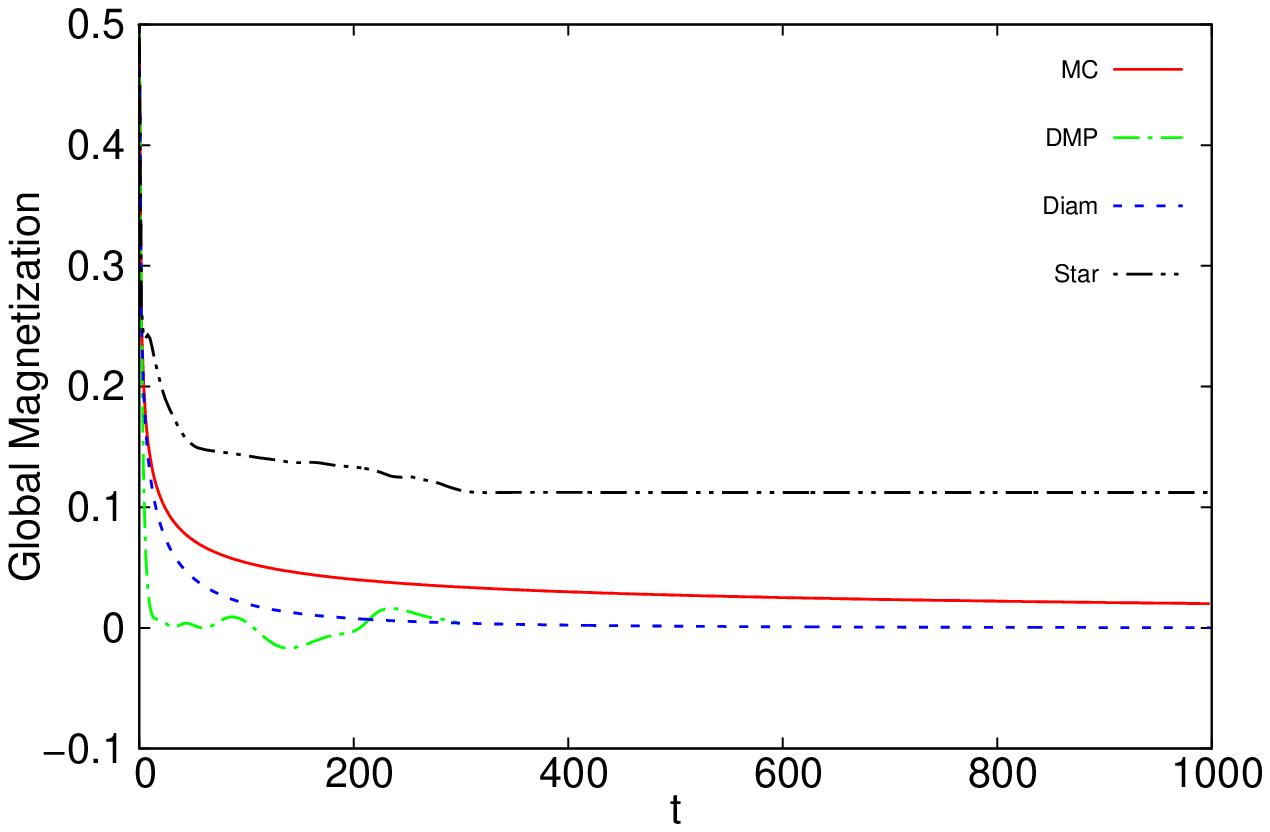}\label{fig:magn_SG_low}}
 \caption{ Spin Glass model with $J_{ij} = \pm 1 /c$. Dynamical behaviour of the magnetization for two different temperatures. Different lines corresponds to MC and the different approximate methods discussed in the main text. Due to the oscillatory behaviour of the parallel dynamics only even times are shown.}
 \label{fig:SG}
\end{figure}

For our dynamical investigation, we study the time evolution of the kinetic Ising spin glass with transition rate defined in \eqref{eq:rate_w} with couplings $J_{ij} = \pm 1/c$ chosen from a bimodal distribution and zero external local field, in the spin glass phase for the temperature $T=0.25$.   
The critical temperature for the spin glass transition is $T_{SG}=0.506$. In Figure \ref{fig:SG} we report the dynamical evolution of the global magnetization for this case obtained by starting from an initial configuration with $m_i= 0.5$ for each site $i$. Due to the parallel dynamic update rule used in this contribution and discussed in Section \ref{sec:kinetic}, both the global and the local magnetization show an oscillatory behaviour for the spin glass case therefore, in Figure \ref{fig:SG} we only show the behaviour for even times. The DMP 1-step Markov memory approach well recovers the transient behaviour only for very short times of the dynamics and then very quickly converges to the equilibrium value of the global magnetization $m=0$. Also the star approximation shows a good accuracy only for very short times and its results for the long time behaviour  are far from MC simulations (as in the other models the star approximation always returns a too large magnetization). 

At variance with these methods, the diamond approximation shows a much better agreement with MC results, both for the short and long time dynamics. In the high temperature regime (see Figure \ref{fig:magn_SG_high}) its results are almost indistinguishable from MC simulations whereas, lowering the temperature, its performances becomes progressively worse for long time dynamics (see Figure \ref{fig:magn_SG_low}).

The decrease of accuracy of the diamond approximation in the SG phase, compared to all the previous cases where it works perfectly, can be understood by the following argument.  Spin glass models are known to have many different \emph{states}, i.e.\ kinds of long range order, in the low temperature phase. Although in each state the local magnetizations are strongly different from zero, their sign change chaotically from state to state; such that, if an average over the states is performed, mean local magnetizations are very close to zero.
So, while the dynamics locally develops one (or few) type of order, having a sensibly non-zero magnetization, the diamond approximation takes the average over all possibile dynamical trajectories and predicts a much smaller magnetization (see right panel in Fig.~\ref{fig:magn_SG_low}).

The CVM-based approximations we are studying here are not designed to take into account the many states present in a spin glass phase; they assume the joint probability distribution can be factorized as in a single thermodynamical pure state. In the spin glass jargon, these approximations are replica symmetric. The replica symmetry can be broken within the CVM framework \cite{rizzo2010replica} and this probably leads to a better approximation for the dynamics in a spin glass phase \cite{lage2014message}. We leave this subject for a future study.

\section{Discussion}
\label{sec:discussion}

In this contribution we have proposed to extend to the non-stationary/transient dynamical regime of Ising-like models the simple variational approach introduced in \cite{kikuchi61} and recently improved in \cite{pelizzola2013variational} so far used to approximate only the stationary regime.

This simple variational formulation is based on two key steps: (i) the construction of a non-equilibrium functional depending on the joint probability distribution of spin histories and (ii) the approximation of this probability according to the prescription given by the cluster variational method. The minimization of the approximated functional, under the constraint of marginalization consistency for the probabilities, leads to simple iterative equations for the joint probabilities of local variables. These iterative equations allow for a computationally very efficient estimation of both macroscopic observables (e.g.\ global magnetization and energy) and microscopic local observables (e.g.\ single spin magnetizations, two-times and two-points correlations).

In \cite{pelizzola2013variational} this approach was shown to give good results for the equilibrium (or in general, stationary) states and to outperform existing methods in the literature.
Here we have tested this approximation for the analytical description of \emph{non-stationary} dynamics of several Ising models defined on a random graph topology: ferromagnets with both symmetric and partially asymmetric couplings, random field models and spin glasses.

The numerical validation has been achieved by a detailed comparison of local microscopic observables (single spin magnetizations and two-times and/or two-spins correlations) with data obtained from extensive Monte Carlo simulations of the dynamics. 
We have found that the \emph{star approximation} in general predicts a too large magnetization: this is probably due to the fact it enforces only self-consistency between single spin magnetizations and, since mean-field approximations tend to stabilize metastable states, the evolution under the star approximation may get easily stuck in a metastable state with a too large magnetization.
On the contrary the \emph{diamond approximation} is extremely accurate in all the models studied even at low temperature. The only situation where it fails to follow the exact dynamical evolution is the low temperature phase of a spin glass model: we believe this is due to the presence of many states, a feature not taken into account by the diamond approximation (which is essentially a replica symmetric approximation in the spin glass jargon).

We have included in the comparison also a method presented in \cite{del2015dynamic} known as dynamic message-passing 1-step Markov memory, that performs in general worst than the diamond approximation. Instead the method presented in \cite{barthel2015matrix} has not been included in the comparison because it is computationally much more demanding; it would be not very fair to compare the goodness of methods that require very different computational resources.

It is worth stressing that the vast majority of the computational time in this work has been dedicated to run a very large number of Monte Carlo simulations to achieve a small error on microscopic observables; the solution of the mean-field equations for star and diamond approximations takes roughly the same running time of a single Monte Carlo trajectory, and the latter approximation outputs mean values for microscopic observables as accurate as Monte Carlo in many cases.
So, in situations where the mean-field approximation is not too crude, the use of the computationally heavy Monte Carlo method can be safely avoided.

The main conclusion is that the diamond approximation is extremely effective in describing the non-stationary regime of the dynamics of Ising models on random graphs.
We think that the prominence of the approach studied here, besides its good results, resides in its simplicity, the intuitive ground from which it is derived, on its cheap computational cost and the possibility of extending it to include high order correlations (ignored in the simplest mean-field approximations). We believe that these features may allow for an easy and immediate application of the method to the investigation of non-equilibrium dynamics of other systems, as for instance biological and social systems.

\FloatBarrier
\textbf{Acknowledgements.}
We thank Erik Aurell for comments and a careful reading of the manuscript. We also acknowledge Alejandro Lage-Castellanos, Roberto Mulet, Alessandro Pelizzola and Marco Pretti for useful comments and discussions. 
This work has been partially supported by the STINT project \textquotedblleft Enhancing cooperation opportunities with Havana University through the Erasmus Programme actions\textquotedblright (ED) and has been funded under FP7/2007-2013/Grant No. 290038 (GDF) and by the European Research Council (ERC) under the European Unions Horizon 2020 research and innovation programme (grant agreement No [694925]).


\appendix

\section{Constrained minimization of $\mathcal{F}$ functional}
\label{app:kikuchi_minimization}
As stated in section \ref{sec:variational_kik_pel}, the dynamic obeyed by a system can be obtained
from a constrained minimization of the $\mathcal{F}[P]$ functional, defined in equation (\ref{eqn:kikuchifreeenergy}).
We will rewrite it here for more clarity:
\begin{eqnarray}
  \mathcal{F}\left[P(\bs^0,\ldots,\bs^{t_f})\right]=\sum_{\bs^0,\ldots,\bs^{t_f}}P(\bs^0,\ldots,\bs^{t_f})
  \left[- \sum_{t=1}^{t_f} \ln W(\bs^{t}| \bs^{t-1})+\ln P(\bs^0,\ldots,\bs^{t_f})\right] 
\end{eqnarray}
The argument of this functional is the joint probability distribution of the histories of all variables. This probability
must be consistent to the initial condition $P(s^0)$ from which the system evolves:
\begin{equation}
\sum_{\bs^1,\ldots,\bs^{t_f}}P(\bs^0,\ldots,\bs^{t_f})=P(\bs^0)
\label{eqn:constraint2}
\end{equation}
The standard procedure to solve this kind of problems is the method of Lagrange multipliers. The Lagrange function in this case reads:
\begin{equation}
 \mathcal{L}[P(\bs^0,\ldots,\bs^{t_f})]=\mathcal{F}[P(\bs^0,\ldots,\bs^{t_f})] - \sum_{\bs^0} \lambda(\bs^0) \left(\sum_{\bs^1,\ldots,\bs^{t_f}}P(\bs^0,\ldots,\bs^{t_f}) - P(\bs^0) \right)
\end{equation}
and the stationary points are the solutions to the system of equations:
\begin{equation}
\begin{cases}
 \dfrac{\partial \mathcal{L}}{\partial \lambda({\bs'}^{0})}=0 & \forall \:\: {\bs'}^{0} \\
 \dfrac{\partial \mathcal{L}}{\partial P({\bs'}^{0},\ldots,{\bs'}^{t_f})}=0 & \forall \:\: ({\bs'}^{0},\ldots,{\bs'}^{t_f})\\
\end{cases}
\label{eqn:minima_condition}
\end{equation}
The first equation in (\ref{eqn:minima_condition}) gives just the constraining relation. 
On the other hand, for the second, it should be noticed that derivatives are taken respect the 
specific value of $P$ for each set of histories $({\bs'}^{0},\ldots,{\bs'}^{t_f})$. After derivation, this second condition leads to:
\begin{equation}
 P({\bs'}^{0},\ldots,{\bs'}^{t_f})=C({\bs'}^{0}) \prod_{t=1}^{t_f} W({\bs'}^{t}| {\bs'}^{t-1}),
\end{equation}
which, using the constraint (\ref{eqn:constraint2}), reduces to an equivalent of the original dynamic of the system (compare to (\ref{eqn:markov1}) and (\ref{eqn:markov2})):
\begin{equation}
 P({\bs'}^{0},\ldots,{\bs'}^{t_f})= \prod_{t=1}^{t_f} W({\bs'}^{t}| {\bs'}^{t-1}) P({\bs'}^{0})
\end{equation}
%

\section{Computation of correlations within the dynamic message passing formalism}\label{app:DMP}

In this Appendix we want to show how to explicitly compute the correlation functions between spin $i$ and its neighbours by using the dynamic message-passing approach presented in \cite{del2015dynamic} in order to reproduce the results illustrated in Section \ref{sec:results}. The computation of these observables is indeed not shown explicitly in \cite{del2015dynamic} although the math ingredients to compute them are all already present there. We therefore believe useful to review these contents in order to clarify how to compute correlation functions within this formalism. The approach is quite general and allows, in principle, to compute correlation functions both at the same time or at different times in the dynamics. Equation (19) of \cite{del2015dynamic}, that we report below, illustrates how to compute the joint probability distribution of spin $i$ and its neighbours at the same time $t$:
\begin{align}\label{eq:iterP} 
P^{(t)}(s_i^t,s_{\partial i}^{t}) &= \sum_{ s_i^{t-1}, s_{\partial i}^{t-1}} \prod_{j\in \partial i}  T_{j \to (ij)} (s_j^{t} | s_j^{t-1}, s_i^{t-1}) \, W_i(s_i^t | s_{\partial i}^{t-1}) \, P^{(t-1)}(s_i^{t-1},s_{\partial i}^{t-1}).
\end{align}
Note that we here adapted the notation of \cite{del2015dynamic} to the symbols adopted in this manuscript. According to \eqref{eq:iterP} the joint probability $P^{(t)}(s_i^t,s_{\partial i}^{t})$ between a given spin and its neighbours at time $t$ can be computed iteratively starting from the initial conditions for $P^{(t)}$ at time zero. Above $W_i(s_i^t | s_{\partial i}^{t-1})$ is the same transition rate for spin $i$ which appears in the main text, whereas $T_{j \to (ij)} (s_j^{t} | s_j^{t-1}, s_i^{t-1})$ represents the two-times message that comes from the $1$-step Markov ansatz taken in \cite{del2015dynamic} and which can be computed also iteratively according to equation (13) in \cite{del2015dynamic}. A partial marginalization of \eqref{eq:iterP} allows then to compute same-time correlations as $c_{ij}^t = \langle s_i^t, s_j^t\rangle$ with $j$ in the neighbourhood of $i$ ($j \in \partial i$), which are though not investigated in this work where we rather focus on correlations at different time. Removing the sum on the RHS of \eqref{eq:iterP} gives the more general two-times joint probability distribution 
\begin{equation} \label{eq:joint_gen}
P^{(t,t-1)}(s_i^t,s_{\partial i}^{t}, s_i^{t-1},s_{\partial i}^{t-1}) = \prod_{j\in \partial i}  T_{j \to (ij)} (s_j^{t} | s_j^{t-1}, s_i^{t-1}) \, W_i(s_i^t | s_{\partial i}^{t-1}) \, P^{(t-1)}(s_i^{t-1},s_{\partial i}^{t-1})
\end{equation}

which can be marginalized to obtain the joint probability function between spin $i$ at time $t$ and its neighbours at time $t-1$ as follows
\begin{equation} \label{eq:joint_twotimes}
P^{(t,t-1)}(s_i^t,\, s_{\partial i}^{t-1}) = \sum_{s_{\partial i}^{t},s_i^{t-1}}\prod_{j\in \partial i}  T_{j \to (ij)} (s_j^{t} | s_j^{t-1}, s_i^{t-1}) \, W_i(s_i^t | s_{\partial i}^{t-1}) \, P^{(t-1)}(s_i^{t-1},s_{\partial i}^{t-1})
\end{equation}

As it is easy to see, the joint probabilities appearing on both sides of \eqref{eq:joint_twotimes} are not the same. For each time of the dynamics the same-time probability on the RHS has indeed to be determined using the iterative equation \eqref{eq:iterP} which can then be plugged into \eqref{eq:joint_twotimes} to obtain the two-times joint probability $P^{(t,t-1)}(s_i^t,s_{\partial i}^{t-1})$. This latter can then be used to compute two-times correlations as $c_{ij}^{(t,t-1)} = \langle s_i^t, s_j^{t-1}\rangle$ with $j$ in the neighbourhood of $i$. This is the procedure used to compute such correlation functions appearing in Section \ref{sec:results}.

\bibliography{bibliography_dynamics}
\end{document}